# First results from cryogenic avalanche detectors based on gas electron multipliers

A. Buzulutskov [*], A. Bondar, L. Shekhtman, R. Snopkov, Y. Tikhonov

*Budker Institute of Nuclear Physics, 630090 Novosibirsk, Russia*

**Abstract**

We study the performance of gaseous and two-phase (liquid-gas) cryogenic detectors of ionizing radiation, based on gas electron multipliers (GEMs). For the first time, high gas gains, exceeding $10^4$, were obtained in pure He, Ar and Kr at low temperatures and the stable avalanche mode of operation was observed in the two-phase cryogenic detector in Kr. The electron avalanching at low temperatures, in the range of 120-300 K, is systematically studied.

*PACs:* 29.40.Cs, 29.90.+r.

## 1. Introduction

Cryogenic detectors of ionizing radiation based on noble gases are widely used in high-energy physics, in particular in noble liquid calorimetry. Practically all such detectors are operated in an ionization mode, i.e. without internal amplification. On the other hand, for scores of experiments with low background physics it would be very attractive to have electron avalanching at low temperatures. However, the attempts to obtain high and stable avalanche amplification at low temperatures had a limited success: either low gains, in noble liquids [1-3], or unstable operation, in two-phase cryogenic detectors [4,5] were observed using wire or needle proportional counters.

In the present paper we show that the problem of electron avalanching at low temperatures, and consequently that of the development of cryogenic avalanche detectors, might be solved using gas electron multipliers (GEMs). The GEM [6] is a thin (50 μm thick) kapton foil metal clad from both sides and perforated by a dense matrix of micro-holes, with a typical hole diameter of 70 μm and inter-hole distance of 140 μm, inside which the gas amplification occurs. It has been shown [7,8] that the multi-GEM structures provide high gain operation in pure noble gases in a wide pressure range, at room temperature. This is a remarkable property, which is unusual for traditional gas avalanche detectors. On the other hand, there is an attractive idea, relevant to two-phase cryogenic detectors being currently developed for dark matter [9] and solar neutrino [10] detection, to use gas avalanche devices operated at cryogenic temperatures for the signal readout. These provided the rationale to begin the present study.

---

[*] Corresponding author. Email: buzulu@inp.nsk.su

In this paper we briefly present first results on the GEM operation at low temperatures in gaseous and two-phase modes. More elaborated presentation will be done elsewhere [11].

## 2. Results

The cryogenic avalanche detector comprised a cryostat chamber cooled with liquid nitrogen, with a triple-GEM amplification structure mounted inside (Fig.1). The GEM electrodes were biased through a high-voltage resistive divider, the bottom of the chamber acting as a cathode. The detector was filled with pure He, Ar or Kr and could operate in either gaseous or two-phase (liquid-gas) mode. In the latter, the ionization produced in the liquid was extracted into the gas phase by an electric field, where it was detected with the help of the multi-GEM multiplier operated in saturated vapor. The detector was irradiated with X-rays, at typical fluxes varying from $10^2$ to $10^5$ $s^{-1}mm^{-2}$, or with β-particles through a thin window at the bottom of the chamber. The liquid layer thickness in the two-phase mode was 3 mm, providing full absorption of the radiation. More detailed description of the experimental setup and procedures will be presented elsewhere [11].

Fig.2 shows the temperature dependence of the triple-GEM gain (gas amplification factor) in He, Ar and Kr at a constant operation voltage and constant gas density, corresponding to a pressure at room temperature of 3, 1 and 1 atm, respectively. The latter conditions allow one to observe fine temperature effects in avalanche mechanism, since in first approximation the gas gain depends only on the electric field and the gas density. The errors shown in this and following figures are systematic.

One can see that in He the gain is practically independent of the temperature, in the range of 120-300 K. This presumably rules out the effect of organic impurities (which might exist for example due to outgassing of kapton and teflon insulation) on the avalanche mechanism in He. Indeed, they would be frozen out at low temperatures. Also, this may speak in support of the hypothesis of associative ionization (the Hornbeck-Molnar process [12]) proposed recently in order to explain the unexpectedly low operation voltages and high gains observed in dense He [13].

In Ar and Kr the gain increased by a factor of 2-3 when decreasing temperature. Obviously, this gain increase cannot be explained in the frame of the electron impact ionization mechanism and is presumably due to the atomic collision-induced ionization, in particular the associative ionization.

Fig.3 shows gain-voltage characteristics of the triple-GEM at low temperatures in gaseous He, Ar and Kr. One can see that rather high gains are reached in all the gases studied. In particular, the maximum gain exceeds $10^5$ and few tens of thousands in He and Ar and Kr, respectively. In addition, we do not observe any unusual properties in the shape of anode pulses, induced just by low temperatures.

It is known that the electron emission from a liquid into the gas phase has a threshold behavior as a function of the electric field due to a potential barrier at the liquid-gas interface [4,5]. This is illustrated in Fig.4 showing the anode current recorded in the cathode gap, induced by X-ray absorption, as a function of the electric field, in Kr. In the gaseous mode the current is independent of the field. In contrast, in the two-phase mode there exist some critical electric field, of about 2 kV/cm: the electron emission from the liquid takes place only when exceeding this field value, in accordance with earlier observations [5]. Therefore, during the measurements in the two-phase mode, the electric field in the cathode gap was kept above the critical value, at about 3 kV/cm.

The gain-voltage characteristic of the triple-GEM in saturated Kr vapor in a two-phase mode is shown in Fig.3. The maximum gain exceeds $10^4$. The detector could operate for at least an hour in the two-phase mode without visible degradation of the gain.

Fig.5 further illustrates the detector performance in a two-phase mode in Kr: anode signals of the triple-GEM are shown at a temperature of 119 K, corresponding to a saturated vapor pressure of 0.94 atm. The signals are induced by β-particles of $^{90}$Sr. From this figure one can roughly estimate the collection efficiency of electrons created in the liquid, taking into account the measured triple-GEM gain, amplifier calibration, average energy deposited in the liquid by β-particle and energy needed for ion pair creation in liquid Kr. The collection efficiency turns out to be of the order of 20%, which is compatible with estimations accounting for the electron life-time in the liquid (at given Kr purity) and electron emission probability at the liquid-gas interface (at given electric field).

## 3. Conclusion

To summarize, we have studied the performance of the cryogenic avalanche detector of ionizing radiation based on triple-GEM multiplier and operated in gaseous and two-phase (liquid-gas) mode in pure He, Ar and Kr. For the first time, high gas gains, exceeding $10^4$, were obtained at low temperatures in noble gases and the stable avalanche mode of operation was observed in the two-phase cryogenic detector in Kr. The electron avalanching has either weak, in He, or moderate, in Ar and Kr, temperature dependence in the range of 120-300 K. Another conclusion is that the GEM structures can successfully operate at low temperatures, down to 120 K. Further studies of this technique are on the way.

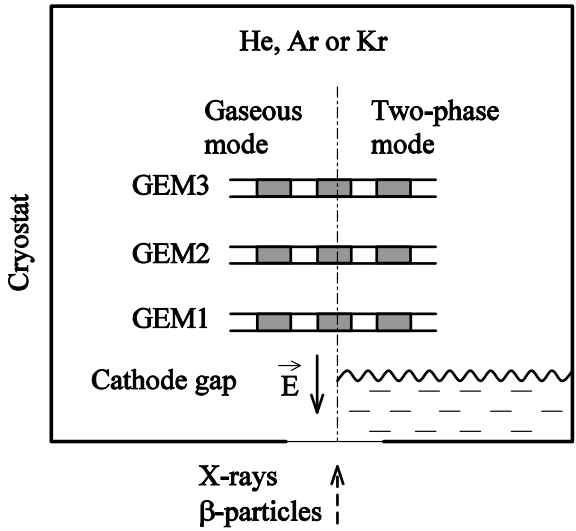

Fig.1 Schematics of the GEM-based cryogenic avalanche detector operated in gaseous and two-phase mode.

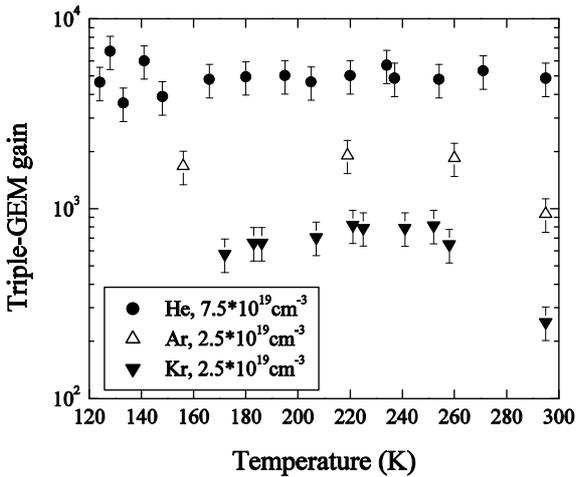

Fig.2 Triple-GEM gain as a function of temperature at a constant operation voltage and constant gas density in He, Ar and Kr. The appropriate atomic densities are indicated.

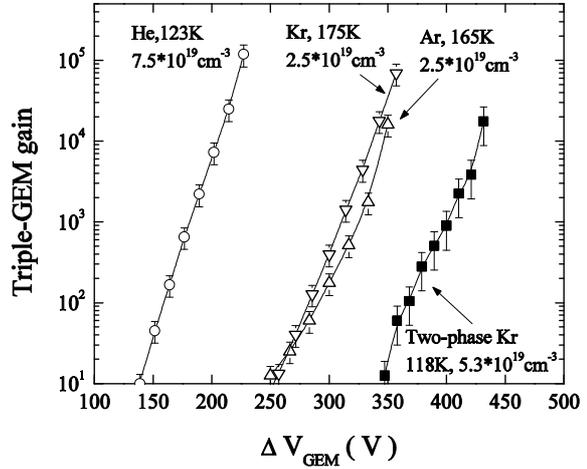

Fig.3 Triple-GEM gain as a function of the voltage applied across each GEM at low temperatures, in gaseous He (at 123 K and 1.25 atm), Ar (at 165 K and 0.56 atm) and Kr (at 175 K and 0.59 atm) and in saturated Kr vapor in two-phase mode (at 118 K and 0.84 atm). The appropriate atomic densities of the gas phase are indicated.

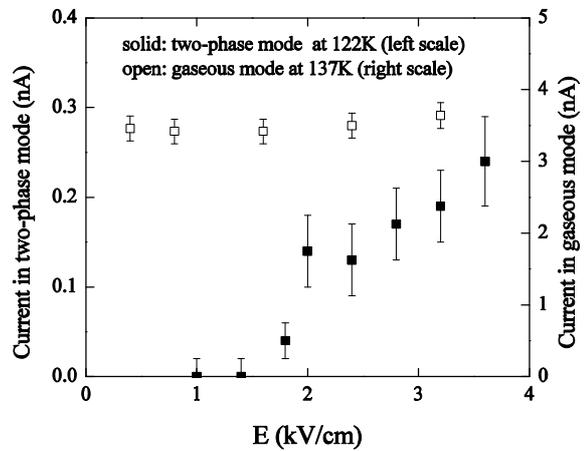

Fig.4 Anode current recorded in the cathode gap as a function of the electric field, in Kr, in two-phase (at 122 K and 1.23 atm) and gaseous (at 137 K and 1.67 atm) mode. The current is induced by X-rays.

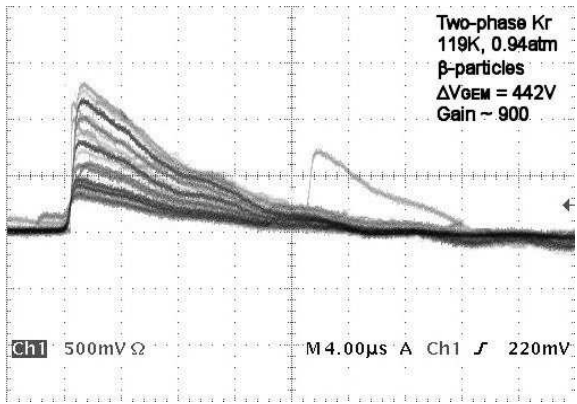

Fig.5 Anode signals from the triple-GEM in two-phase Kr, after a charge-sensitive amplifier, induced by β-particles of $^{90}$Sr and obtained at a triple-GEM gain of 900, temperature of 119 K and pressure of 0.94 atm.